\documentclass[prd,aps]{revtex4-1}

\usepackage{epsfig}
\usepackage{graphicx,color}

\newcommand{\beq}{\begin{equation}}
\newcommand{\eeq}{\end{equation}}
\newcommand{\be}{\begin{eqnarray}}
\newcommand{\ee}{\end{eqnarray}}

\begin{document}

\title{
Lifetime of the dark $Z$ boson
}

\author{Dong-Won Jung}
\email{dongwon.jung@yonsei.ac.kr}
\affiliation{Department of Physics, 
Yonsei University, Seoul 03722, Korea}
\affiliation{Department of Physics,
Chonnam National University, Gwangju 61186, Korea}

\author{Kang Young Lee}
\email{kylee.phys@gnu.ac.kr, Corresponding author.}

\author{Chaehyun Yu}
\email{chyu@kias.re.kr}


\affiliation{ Department of Physics Education and RINS,
Gyeongsang National University, Jinju 52828, Korea}

\date{\today}

\begin{abstract}

The mediator particle
between the Standard Model sector and a hidden sector 
might have a long lifetime to show 
the observable displaced vertices in experiments.
Considering a fermionic dark matter model in which the hidden sector is
connected to the Standard Model 
by an additional Higgs doublet field,
the mediator dark $Z$ boson may live long enough.
We explore the possibility to observe
the displaced vertices of the long-lived dark $Z$ boson
at the CERN LHC
and at the proposed SHiP experiment.
We find that the ATLAS and CMS searches 
for the long-lived dark $Z$ boson
can probe the mass range 
$7 < m_{Z'} < 150~{\rm MeV}$
with 150 fb$^{-1}$ integrated luminosity at the LHC run 3,
and the SHiP experiment will probe
$2 m_e < m_{Z'} < 15~{\rm MeV}$ range
with $6 \times 10^{20}$ protons on target
in total 15 years.
The dark matter phenomenology is also discussed
in the region where such a long-lived mediator is detectable.

\end{abstract}

\pacs{ }

\maketitle

\section{Introduction}

Search for new long-lived particles (LLPs)
is one of the most exciting issues 
of the current particle physics experiments
\cite{Alimena:2019zri,Knapen:2022afb}.
The LLPs travel substantial distances 
from the point where they are created
before decaying.
The experimental signatures of LLPs might be
very different from those of the Standard Model (SM) processes.
If the LLPs decay into the SM particles inside the detector, 
displaced vertices of multijets or lepton pairs are created
and we can observe exotic signatures
of new physics beyond the SM.

We have suggested a hidden sector model
where the hidden sector consists of a singlet fermion 
with the hidden U(1)$_X$ gauge symmetry
\cite{Jung:2020ukk}.
The singlet fermion in the hidden sector
is a dark matter (DM) candidate.
We introduce an additional Higgs doublet 
as a mediator field between
the hidden sector and the SM sector.
The additional Higgs field has the hidden U(1)$_X$ charge
and does not couple to the SM fermions
because of the U(1)$_X$ charge.
The U(1)$_X$ gauge symmetry is spontaneously broken
by the electroweak symmetry breaking (EWSB) in this model.
There exists an extra massive gauge boson $Z'$
mixed with the SM $Z$ boson as a result.
We call it a dark $Z$ boson
since its couplings to the SM fermions are
proportional to those of the $Z$ boson
up to the suppression factor of the mixing angle.
The dark $Z$ boson mass is of order the electroweak scale,
and could be very light in this model.
Thus the dark $Z$ boson gets stringent constraints
from the electroweak processes 
at colliders and the low energy neutral current (NC) experiments.
Moreover, masses of the additional Higgs bosons can be light 
and the Higgs sector phenomenology strongly limits the model parameters
\cite{Jung:2020ukk,Jung:2021bjt,Jung:2023ckm}.
As a result of the electroweak and Higgs sector constraints,
the light dark $Z$ boson requires very small couplings
which lead to the long lifetime.

In this work, we explore the possibility 
to detect the dark $Z$ boson 
by observing the displaced vertices 
in the present and future experiments.
There have been various attempts to search for the LLPs
in the existing experiments
ATLAS \cite{ATLAS:2022izj}
and CMS \cite{CMS:2021kdm}
at the CERN LHC,
and in the proposed experiments
SHiP \cite{SHiP,Choi:2016vic,SHiP-2023proposal},
MATHUSLA \cite{MATHUSLA},
and newly operating LHC far detectors
SND@LHC \cite{SND,Choi:2022qzk} and FASER \cite{FASER}.

The ATLAS and CMS detectors are looking for 
target signatures of the LLPs 
using the inner tracking detector
\cite{ATLAS:2019jcm,ATLAS:2019tkk,CMS:2023gpb,CMS:2022qej,CMS:2021tkn},
calorimeters 
\cite{ATLAS:2022zhj}
and muon spectrometer 
\cite{ATLAS:2022gbw,CMS:2023arc}.
In order to make an observed vertex,
the dark $Z$ mass should be larger than
2 times the electron mass at least.
On the other hand, 
the dark $Z$ promptly decays and leaves no displaced vertex
if $m_{Z'} > 1$ GeV.
The lifetime of the dark $Z$ 
to be observed inside the LHC detectors is 
of order $\sim {\cal O}$(ns).
Hence the relevant mass range of the dark $Z$ boson 
is from a few MeV to a few hundreds of MeV
and the $Z'$ decay channel is only an electron pair 
for most values of $m_{Z'}$.
Therefore the searches at the inner tracker is essential
and the inner tracker size, $< 1$ m,
stipulates the limit of the $Z'$ searches.

The newly proposed SHiP (Search for Hidden Particles) 
experiment at the CERN SPS
is the dedicated experiment to search for the LLPs.
The SPS accelerator will deliver very high intensity proton beam
$\sim 4 \times 10^{19}$ protons on target per year
for the SHiP experiment.
The SHiP experiment is designed to detect 
decay signatures of the LLPs directly
by full reconstruction and particle identification
of the SM final states.
Since the SHiP facility consists of 
the long vacuum vessel of 50 m length 
located downstream far from the target, more than 30 m,
it will investigate different region of the dark $Z$ mass 
from the LHC and 
play a complementary role for probing the dark $Z$ boson.

We will investigate the lifetime of the dark $Z$ boson
and study the discovery potential of the long-lived $Z'$ boson
at the LHC run 3 and the future SHiP experiments.
This paper is organized as follows.
The model is described 
together with the electroweak and the Higgs sector constraints
to achieve the allowed parameter set in Sec.~2. 
The lifetime and decay length of the dark $Z$ boson 
are discussed in Sec.~3.
The mass ranges of the dark $Z$ boson 
to be observed at the LHC and proposed SHiP experiment
are explored in Sec.~4.
We show that our DM candidate is acceptable
for the relic density of the universe
and the present direct detection experiments
in Sec.~5.
Section 6 is devoted to the summary of the results and conclusion.

\section{The Higgs doublet mediator and the dark $Z$ boson}

We consider a hidden sector
which consists of a Dirac fermion with U(1)$_X$ gauge symmetry.
The hidden sector fermion is a SM gauge singlet and
assumed to be a dark matter candidate. 
The U(1)$_X$ does not acts on the SM fields 
and the hidden sector is not coupled to the SM sector directly.
We assume that the kinetic mixing of U(1)$_X$ 
with the SM U(1)$_Y$ is ignored.
Thus we introduce an extra Higgs doublet $H_1$ as
a mediator field between the SM and the hidden sector.
The gauge charges of the hidden sector fermion $\psi$
and the mediator $H_1$ are assigned to be
\be
\psi (1, 1, 0, X), ~~~~
H_1 \left(1, 2, \frac{1}{2}, \frac{1}{2} \right),
\ee
based on the gauge group 
${\rm SU}(3)_c \times {\rm SU}(2)_L \times {\rm U}(1)_Y 
\times {\rm U}(1)_X$.
The SM sector also includes a SM-like Higgs doublet $H_2$
with the charge assignment $H_2 (1, 2, 1/2, 0)$.

The Lagrangian of the model is given by
\be
{\cal L} = {\cal L}_{\rm SM-H} + {\cal L}_{\rm hidden} + {\cal L}_H,
\ee
where ${\cal L}_{\rm SM-H}$ is the SM Lagrangian without the Higgs sector.
The hidden sector Lagrangian is given by
\be
{\cal L}_{\rm hidden} = -\frac{1}{4} F_X^{\mu \nu} F_{X \mu \nu} 
                    + \bar{\psi}_X i \gamma^\mu D_\mu \psi_X
                          - m_X \bar{\psi}_X \psi_X,
\ee
where the covariant derivative is given by
\be
D^\mu = \partial^\mu + i g_s G^{\mu ~a} \lambda^a
                 + i g W^{\mu~i} T^i
                 + i g' B^\mu Y + i g_X A_X^\mu X.
\ee
Independent of the structure of the visible sector,
two additional parameters, 
the U(1)$_X$ charge $X$ and the hidden fermion mass $m_X$,
are introduced in the hidden sector.
We also have the Higgs sector Lagrangian
\be
{\cal L}_H = (D^\mu H_1)^\dagger D_\mu H_1 
	   + (D^\mu H_2)^\dagger D_\mu H_2 - V(H_1, H_2) 
	   + {\cal L}_{\rm Y}(H_2), 
\ee
where the Higgs potential
\be
V(H_1,H_2) &=& \mu_1^2 H_1^\dagger H_1 + \mu_2^2 H_2^\dagger H_2
\nonumber \\
	   && + \lambda_1 (H_1^\dagger H_1)^2
	      + \lambda_2 (H_2^\dagger H_2)^2
              + \lambda_3 (H_1^\dagger H_1)(H_2^\dagger H_2)
              + \lambda_4 (H_1^\dagger H_2)(H_2^\dagger H_1),
\ee
and the Yukawa interactions,
\be
-{\cal L}_Y = g_{ij}^d \bar{Q}^i_L H_2 d^j_R
              + g_{ij}^u \bar{Q}^i_L \tilde{H}_2 u^j_R
              + g_{ij}^l \bar{L}^i_L H_2 l^j_R   + H.C. ,
\ee
with $\tilde{H}_2 = i \sigma_2 H^*$.
Prevented are
the soft breaking quadratic term and
the $\lambda_5$ term in the potential 
and the mediator $H_1$ couplings to the SM fermions 
in the Yukawa terms
by the U(1)$_X$ charge.
Thus the flavour structure of our model is 
the same as that of the 2HDM of type I
without a discrete symmetry.
Such a Yukawa structure of the 2HDM of type I
has been discussed in Ref.~\cite{Ko:2013zsa}.
The hidden sector is connected to the SM sector
through the $H_1$
with the SU(2) charge of $H_1$ 
and the Higgs quartic interactions in the potential.

The EWSB is achieved by evolving two vacuum expectation values (VEVs),
$ \langle H_i \rangle = v_i/\sqrt{2} $ 
and the gauge boson mass terms are generated.
We require $v_1^2 + v_2^2 = v^2$, 
with the electroweak scale $v=246$ GeV,
and define $\tan \beta \equiv v_2/v_1$.
Diagonalizing the gauge boson mass matrix, we get masses 
\be
m_{Z,Z'}^2 = \frac{1}{8} \left(
          g_X^2 v_1^2 + (g^2 + {g'}^2) v^2 \pm
          \sqrt{(g_X^2 v_1^2 - (g^2 + {g'}^2) v^2)^2
               + 4 g_X^2 (g^2 + {g'}^2) v_1^4} \right),
\ee
the mixing angle $\theta_X$
\beq
\tan 2 \theta_X = \frac{-2 g_X \sqrt{g^2 + {g'}^2} v^2 \cos^2 \beta}
                       {(g^2+{g'}^2)v^2 - g_X^2 v^2 \cos^2 \beta},
\eeq
and physical states
\be
A_X &=& c_X Z' + s_X Z,
\nonumber \\
W^3 &=& -s_X c_W Z' + c_X c_W Z + s_W A,
\nonumber \\
B &=& s_X s_W Z' - c_X s_W Z + c_W A,
\ee
where $s_X = \sin \theta_X$, $c_X = \cos \theta_X$,
and $s_W = \sin \theta_W$, $c_W = \cos \theta_W$,
with the Weinberg angle $\theta_W$.
The new parameters in the electroweak sector
are the U(1)$_X$ gauge coupling $g_X$ and 
the VEV ratio $\tan \beta$. 
Instead of them, we present the analysis in terms of
the dark $Z$ mass $m_{Z'}$ and the mixing angle $s_X$ 
expressed by
\be
s_X^2 = \frac{m_W^2 \cos^2 \beta}
         {c_W^2 (m_Z^2-m_{Z'}^2)-m_W^2 \cos^2 \beta}
        \frac{m_{Z'}^2}{m_Z^2-m_{Z'}^2},
\ee
hereafter.
We write the neutral current interactions
in terms of the physical gauge bosons as
\be
{\cal L}_{NC} 
= - e A^\mu \bar{f} Q \gamma_\mu f
    - c_X Z^\mu \left( g_V \bar{f} \gamma_\mu f
          + g_A \bar{f} \gamma_\mu \gamma_5 f \right)
    + s_X {Z'}^\mu \left( g_V \bar{f} \gamma_\mu f
          + g_A \bar{f} \gamma_\mu \gamma_5 f \right),
\ee
where the couplings $e$, $g_V$, $g_A$ are
defined in the same manner as those of the SM.
Note that the $Z'$ couplings are the same as the ordinary $Z$ couplings
up to the suppression factor $\tan \theta_X $,
which is the reason why we call the extra gauge boson $Z'$ the dark $Z$.

We apply the constraints from the electroweak processes
to the model parameters.
The $Z$ boson mass is shifted by the $Z$--$Z'$ mixing
in this model but the $W$ boson mass is not.
Thus the $\rho$ parameter defined by the ratio of 
$W$ and $Z$ boson masses, 
$ \rho \equiv m_W^2/m_Z^2 c_W^2$, is modified.
Moreover, there also exist the new scalar contributions 
to the $\rho$ parameter at loop levels.
We write the deviation of the $\rho$ parameter 
from the SM value as
$\Delta \rho= \Delta \rho_X + \Delta \rho_H$, where 
\be
\Delta \rho_X \equiv 1 - \frac{1}{\rho}
     \approx -s_X^2 \left(1 - \frac{m_{Z'}^2 c_W^2}{m_W^2} \right),
\ee
in the leading order of $s_X$.
The scalar loop contribution $\Delta \rho_H$ 
can be found in Ref.~\cite{Jung:2023ckm}.
The measured value of $\Delta \rho$ is obtained from 
the $T$ variable by the relation
$\Delta \rho = \alpha(m_Z) T$
with $T = 0.09 \pm 0.07 $,
and the fine structure constant at five loop level
$\alpha^{(5)}(m_Z)^{-1} = 127.955 \pm 0.010$
\cite{PDG}. 

Next we consider the atomic parity violation (APV) 
in the precise measurement of the weak charge of nuclei.
Including the dark $Z$ contribution, 
the weak charge is given by
\beq
Q_W = Q_W^{SM} \left( 1+\frac{m_Z^2}{m_{Z'}^2} s_X^2 \right),
\eeq
in the leading order of $s_X$.
Using the SM value \cite{APVSM1,APVSM2}
$Q_W^{SM} =  -73.16\pm0.03$,
the present experimental value for the Cs atom
given by \cite{APV} $ Q_W^{exp} = -72.82\pm0.42$,
yields the constraint on this model
\beq
\frac{m_Z^2}{m_{Z'}^2} s_X^2 \le 0.006,
\eeq
at 95 \% C.L \cite{hslee}.

\begin{figure}[t]
\centering
\includegraphics[height=7cm]{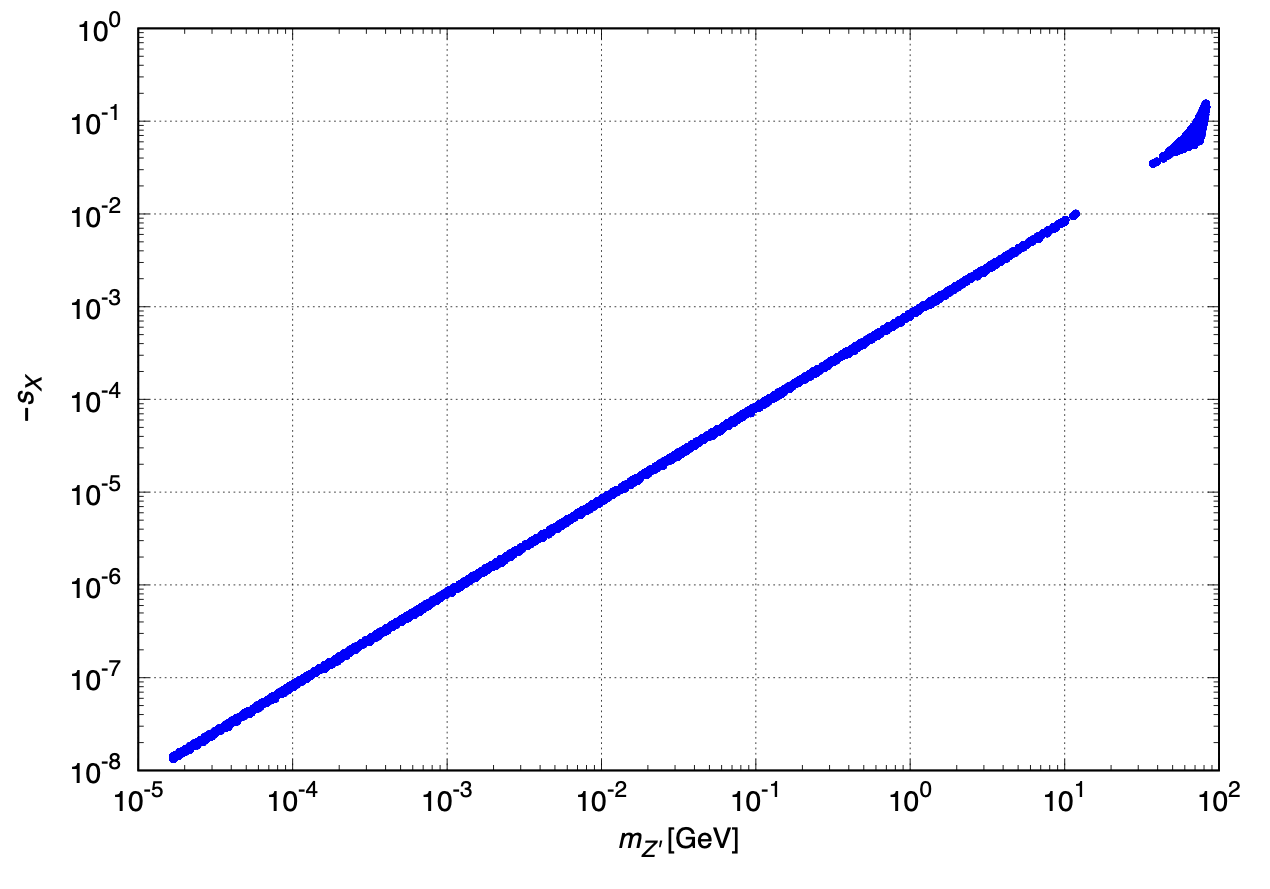}
\caption{
Allowed values of $(m_{Z'}, |\sin \theta_X|)$
by the electroweak and Higgs phenomenology constraints~\cite{Jung:2023ckm}.
} 
\end{figure}

After the EWSB, we diagonalize the mass matrices 
of the neutral and charged Higgs bosons 
to derive physical states.
There are two CP-even neutral Higgs bosons
and a pair of the charged Higgs boson.
The SM-like Higgs boson, $H$ has the mass of the measured value,
$ m_H = 125.10$ GeV,
and the other neutral Higgs boson, $h$ has the mass $m_h$ 
which is a free parameter.
The charged Higgs boson mass $m_\pm^2 = - \lambda_4 v^2/2$
is also a free parameter.
We write the mixing angles, $\alpha$ between neutral Higgs bosons and 
$\beta$ between charged Higgs bosons,
which are also independent model parameters.

For phenomenological analyses,
we require some theoretical constraints on the Higgs sector.
The Higgs quartic couplings should be perturbative,
$|\lambda_i|<4 \pi$.
The Higgs potential should be bounded from below
\be
\lambda_1 > 0,~~~~~
\lambda_2 > 0,~~~~~
\lambda_3 > -2\sqrt{\lambda_1 \lambda_2},~~~~~
\lambda_3 + \lambda_4 > -2\sqrt{\lambda_1 \lambda_2}.
\ee
We also insist the perturbative unitarity 
of the $WW$ scattering at tree level. 

The experimental limits on the Higgs sector
are applied to the model with huge data 
provided by collider experiments, at LEP, Tevatron and LHC.
We use the public codes 
{\tt HiggsBounds} \cite{Bechtle:2020pkv}
which compares various predictions in the Higgs sector
with experimental data 
and use {\tt HiggsSignals} \cite{Bechtle:2020uwn}
which performs a $\chi^2$ test of the Higgs sector predictions
against the signal rate and mass measurements from colliders.
{\tt HiggsBounds} provides the exclusion limits
at the 95$\%$ C.L. for the extended Higgs boson searches.
{\tt HiggsSignals} evaluates the P-value which is demanded
to be less than $0.05$.

Additionally we consider the $Z$ total width 
by inclusion of the $Z \to Z' h$ decay.
Since the dark $Z$ boson and the additional neutral Higgs $h$
can be light in this model,
the new decay channel of the $Z$ boson, $Z \to Z' h$ arises
depending on the sum of the masses of the dark $Z$ and $h$.
The dark $Z$ decays are suppressed 
by the $Z$-$Z'$ mixing $\sin \theta_X$
and the $h$ decays suppressed by $\sin \alpha$.
Thus the $Z \to Z' h$ decay may be an invisible decay
or present characteristic displaced vertices.
The $Z$ total width $\Gamma_Z$ is precisely measured at LEP and SLD
\cite{ALEPH:2005ab}, $ \Gamma_Z = 2.4955 \pm 0.0023 ~~{\rm GeV}$,
and shows a good agreement with the SM prediction,
$\Gamma_Z^{\rm SM} = 2.4941 \pm 0.0009$ GeV.
Hence the new decay channel of the $Z$ boson is very limited
and presents a strong constraints on the NP effects.
By adding the partial width of the $Z \to Z' h$ decay 
to the SM total width,
$\Gamma_Z^{\rm New} = \Gamma_Z^{\rm SM} c_X^2 + \Gamma(Z \to Z' h)$,
and we constrain the model prediction by the experimental limits.

We summarize the independent model parameters used in this work: 
\be
(m_{Z'}, \sin \alpha, \tan \beta, m_{h}, m_\pm ),
\label{parameters}
\ee
while we will also use the $Z$-$Z'$ mixing angle $\theta_X$ 
given in Eq. (11)
for convenience of the phenomenological analysis. 
Scanning the free parameters of Eq.~(\ref{parameters})
with the electroweak and Higgs constraints altogether,
we obtain the allowed parameter set and show them on the
$(m_{Z'},|\sin \theta_X|)$ plane in Fig.~1,
which is necessary for analysis of this work.
Allowed parameter sets for other parameters can be also found 
in Ref.~\cite{Jung:2023ckm}.

\section{Lifetime of the dark $Z$ boson}

Being kinematically allowed,
the dark $Z$ boson can also decay into the DM fermion pair.
The decay width of $Z'$ into the DM fermion pair is given by
\be
\Gamma(Z' \to \bar{\psi}_X \psi_X)
=\frac{(g_X X)^2 m_{Z'}}{12 \pi}
\sqrt{1-4{r}_\psi} \left( 1+2 {r}_\psi \right),
\ee
where ${r}_\psi = m^2_{\psi_X}/m^2_{Z'}$ 
and $X$ is the U(1)$_X$ charge of $\psi_X$.
The decay width of $Z'$, and consequently the lifetime of $Z'$
depend on the free parameter $X$.
On the other hand, 
the dark $Z$ boson decays into the SM particles are
derived from the ordinary $Z$ boson decays,
\be
\Gamma(Z' \to f \bar{f}) = \Gamma(Z \to f \bar{f}) 
   \left( \frac{s_X}{c_X} \right)^2 \frac{m_{Z'}}{m_Z} F(r_f),
\ee
where
\be
F(r) = \sqrt{1-4r} \left[ 1-\frac{8}{3} s_W^2 + \frac{32}{9} s_W^4
      -r ~\left( 1+ \frac{16}{3} s_W^2-\frac{64}{9} s_W^4 
       \right) \right],
\ee
with $r_f = m_f^2/m_{Z'}^2$.
We obtain the ratio of decay widths, $\Gamma(Z' \to f \bar{f})
/\Gamma(Z' \to \bar{\psi}_X \psi_X) \sim |s_X/g_X X|^2 $.
With the typical size of the gauge coupling, 
$g_X X \sim 10^{-2} - 10^{-1}$,
this ratio is very small if $m_{Z'} < 1$ GeV
and the dark $Z$ boson dominantly decays into the DM fermion pair.
Then this decay is invisible and not our concern.
Thus we assume that the dark $Z$ boson
cannot decay into DM by putting $m_{Z'} < 2 m_{X}$ hereafter.

\begin{figure}[t]
\centering
\includegraphics[width=10cm]{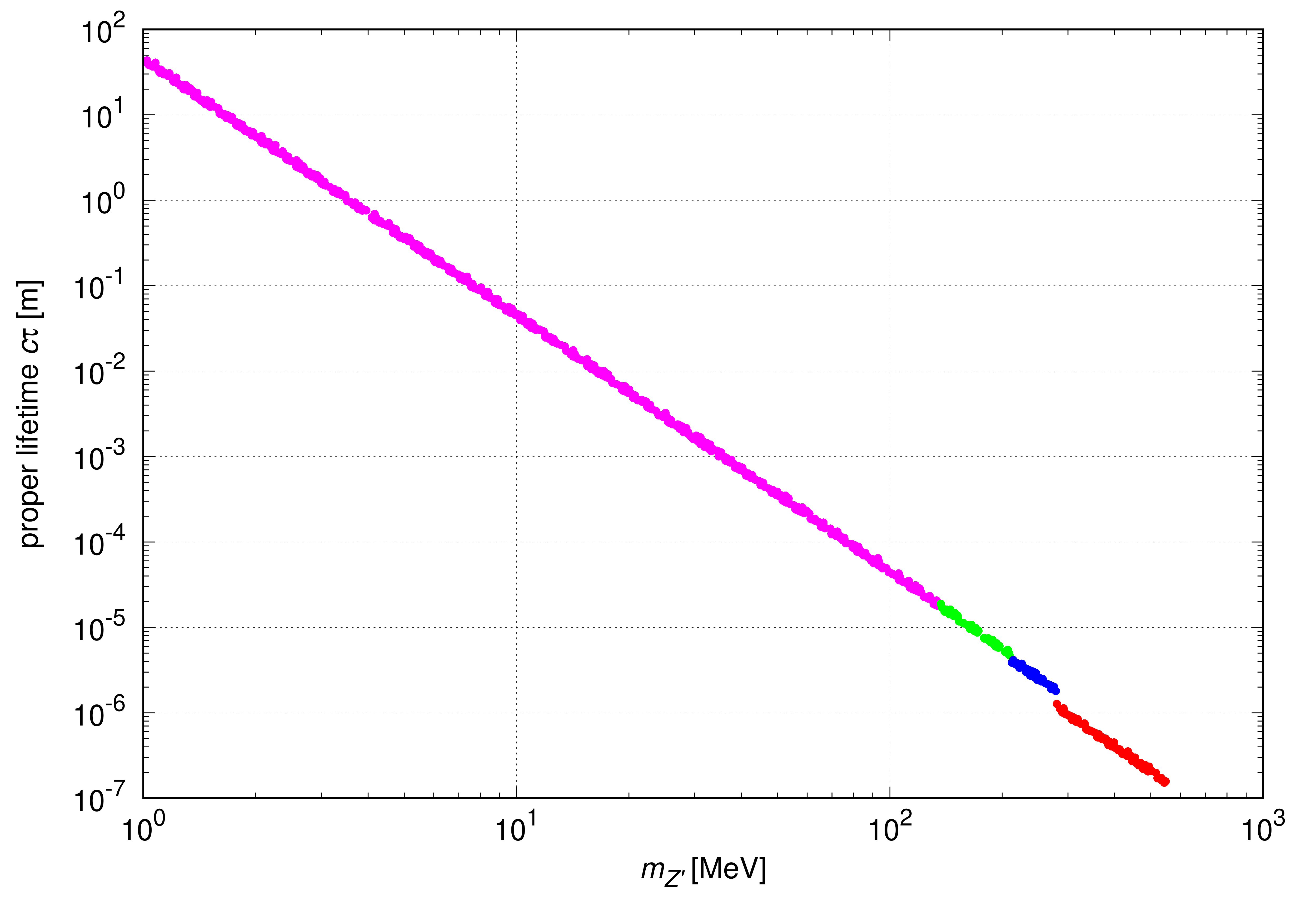}
\caption{
The mean proper lifetime of the dark $Z$ boson $c \tau$
with respect to the mass.
Each color from the left to right denotes the $m_{Z'}$ region, 
$2~m_e < m_{Z'} < m_{\pi^0}$, 
$m_{\pi^0} < m_{Z'} < 2~m_{\mu}$, 
$2~m_\mu < m_{Z'} < 2 m_\pi$,
$2~m_{\pi} < m_{Z'} <  m_{\eta}$, respectively. 
}
\end{figure}

\begin{table}[b] 
\begin{tabular}{cccccc} 
\hline
\hline
$m_{Z'}$ & $|\sin \theta_X|$ & $\Gamma_{Z'}^{\rm total}$ (GeV) & $\tau_{Z'}$ (s) & $\gamma c \tau_{Z'}$ (m) & $E_{Z'}$ (GeV)
\\
\hline
1.5 MeV & $\sim 10^{-6}$ & $9.8 \times 10^{-18}$ & $0.7 \times 10^{-7}$ & $1.4 \times 10^4$ & 1
\\
10 MeV & $\sim 10^{-5}$ & $6.5 \times 10^{-15}$ & $1.0 \times 10^{-10}$ & $3.0$ & 1
\\
100 MeV & $\sim 10^{-4}$ & $6.5 \times 10^{-12}$ & $1.0 \times 10^{-13}$ & $3.0 \times 10^{-4}$ & 1
\\
\hline
1.5 MeV & $\sim 10^{-6}$ & $9.8 \times 10^{-18}$ & $0.7 \times 10^{-7}$ & $1.4 \times 10^5$ & 10
\\
10 MeV & $\sim 10^{-5}$ & $6.5 \times 10^{-15}$ & $1.0 \times 10^{-10}$ & $30$ & 10
\\
100 MeV & $\sim 10^{-4}$ & $6.5 \times 10^{-12}$ & $1.0 \times 10^{-13}$ & $3.0 \times 10^{-3}$ & 10
\\
\hline
250 MeV & $\sim 2 \times 10^{-4}$ & $7.6 \times 10^{-11}$ & $8.7 \times 10^{-15}$ & $1.0 \times 10^{-4}$ & 10
\\
500 MeV & $\sim 5 \times 10^{-4}$ & $2.4 \times 10^{-9}$ & $2.7 \times 10^{-16}$ & $1.6 \times 10^{-6}$ & 10
\\
2 GeV & $\sim 1 \times 10^{-3}$ & $3.9 \times 10^{-8}$ & $1.7 \times 10^{-17}$ & $2.6 \times 10^{-8}$ & 10
\\
\hline
250 MeV & $\sim 2 \times 10^{-4}$ & $7.6 \times 10^{-11}$ & $8.7 \times 10^{-15}$ & $1.0 \times 10^{-3}$ & 100
\\
500 MeV & $\sim 5 \times 10^{-4}$ & $2.4 \times 10^{-9}$ & $2.7 \times 10^{-16}$ & $1.6 \times 10^{-5}$ & 100
\\
2 GeV & $\sim 1 \times 10^{-3}$ & $3.9 \times 10^{-8}$ & $1.7 \times 10^{-17}$ & $2.6 \times 10^{-7}$ & 100
\\
\hline
\hline
\end{tabular}
\caption{
Lifetimes and decay lengths for benchmarking values of 
$m_{Z'}$ and $E_{Z'}$.
}
\label{tab1}
\end{table}

Generically the decay widths of the dark $Z$
into the SM particles are suppressed
by the mixing angle $s_X^2$ and the mass ratio $m_{Z'}/m_Z$.
Since the allowed $s_X$ values are heavily influenced 
by $m_{Z'}$ as shown in Fig.~1,
the decay rate of $Z'$ is almost determined by $m_{Z'}$.
Figure 1 also tells us that smaller the mass, smaller the mixing angle
and the decay width rapidly decreases
as the $Z'$ mass becomes smaller.

The opening channels of the $Z'$ decays rely on the $Z'$ mass.
If $m_{Z'} < 2 m_e$,
the dark $Z$ boson decays into neutrinos only.
Although these decay modes are invisible,
we estimate the lifetime here for comparison of other cases.
If $m_{Z'} \sim 1$ MeV,
$|s_X| \sim 10^{-7}$ from Fig.~1.
With the SM value $\Gamma(Z' \to \nu \bar{\nu})=0.17$ GeV at tree level,
we obtain the width $\Gamma_{Z'} \sim 5 \times 10^{-20}$ GeV
and the lifetime 
$ \tau_{Z'} \sim \hbar/\Gamma_{Z'}^{\rm total} \sim 10^{-5}~{\rm s}$.
Assuming $E_{Z'} = 1$ GeV,
we find that the decay length of the $Z'$ is estimated to be
\be
\gamma c \tau = \frac{E_{Z'}}{m_{Z'}} \cdot 3 \times 10^8 \cdot 10^{-5}
              \sim 3 \times 10^{6} ~({\rm m}),
\ee
and it will escape from the detector.

If $2 m_e < m_{Z'} < 2 m_\mu$, 
the decay channel into electron pair opens and
the decay rate is
\be
\Gamma_{Z'}^{\rm total} 
   = ( 3 \times \Gamma(Z \to \nu \bar{\nu}) + \Gamma(Z \to e^- e^+) F(r_e))
   \left( \frac{s_X}{c_X} \right)^2 \frac{m_{Z'}}{m_Z}
	 = ( 3 \times 0.17 + 0.085 F(r_e))
   \left( \frac{s_X}{c_X} \right)^2 \frac{m_{Z'}}{m_Z}.
\ee
We find that the dark $Z$ boson of
a few tens of MeV will show decay length of ${\cal O}$(cm-m)
and could decay inside the detector 
so that we can observe the displaced vertex. 

When $2m_\mu < m_{Z'} < 2m_\pi$,
the decay channel into the muon pair opens.
The sum of the decay widths of the ordinary $Z$ boson
into the neutrinos, electron and muon pairs are
$\Gamma(Z \to 3 \nu \bar{\nu})+ \Gamma(Z \to e^- e^+)
+ \Gamma(Z \to\mu^- \mu^+) = 0.68$ GeV.
When $2 m_\pi < m_{Z'} < 2 m_c$,
many light hadron channels open from the light quark pairs,
the decay width results in
$\Gamma(Z \to 3 \nu \bar{\nu})+ \Gamma(Z \to e^- e^+)
+ \Gamma(Z \to\mu^- \mu^+)
+ \Sigma_{q=u,d,s} \Gamma(Z \to q\bar{q}) = 1.76$ GeV.
We show the proper lifetime of the dark $Z$ boson 
with respect to its mass in the Fig.~2.

In order to find the observable mass range of the dark $Z$,
we consider a few benchmark points of $m_{Z'}$ 
and present the corresponding lifetimes and decay lengths 
in the Table \ref{tab1}.
We assume the energy of $Z'$ to be 1,10 GeV and 100 GeV
to get the decay lengths.

\section{Search for Long-Lived dark $Z'$ in the experiments}

One of the characteristic signatures of the neutral LLPs is
the displacement of the vertex from the interaction point.
The decay length of the LLP depends on
the energy as well as the lifetime of the produced particle.
If the decay length is too short, 
we cannot observe the displaced vertex,
while too long decay length results in an escape of the particle
from the detector.

In order to observe a displaced vertex 
inside the detector at the LHC,
we need the decay length between ${\cal O}$(mm) and ${\cal O}$(m).
For the SHiP experiment, 
the relevant decay length is ${\cal O}(10-100)$ m.
The decay probability $P_{\rm decay}$ is defined by the probability
that the dark $Z$ boson decays inside the decay volume,
\be
P_{\rm decay} = \exp \left( -\frac{l_i}{l_{\rm decay}} \right) 
	       - \exp \left( -\frac{l_f}{l_{\rm decay}} \right) ,
\ee
where $l_i$ and $l_f$ 
are the minimum and maximum distances
from the creation point of the dark $Z$ 
to the point that the displaced vertex can be observed
and $l_{\rm decay} = \gamma c \tau_{Z'}$ 
is the decay length of the dark $Z$ boson.

Generically the dark $Z$ boson can be produced 
at accelerator experiments in three ways;
the meson decays, proton bremsstrahlung,
and direct perturbative QCD processes.
Among them, a light $Z'$, $m_{Z'}<500$ MeV,
would be mostly produced through meson decays.
As is estimated in the previous section,
the mass range of $Z'$ 
which has decay lengths probed at the LHC or the SHiP
is ${\cal O} (1-100)$ MeV.
Hence we forecast that the dark $Z$ boson
will be produced from meson decays in this work.

The principal source of $Z'$ production 
would be the $\pi^0$ decay if $m_{Z'}< m_{\pi^0}$
since it is the most produced meson 
at high energy accelerator experiments.
The $\pi^0$ meson dominantly decays into 2 photons,
and the $\pi^0 \to \gamma \gamma$ decay is described 
by the chiral Lagrangian method.
We can obtain the $\pi^0 \to Z' \gamma$ decay rate
by replacing one $\gamma$ into a dark $Z$ boson
with changing couplings
$\alpha^2 \to \alpha \alpha_{Z'}$,
where $\alpha_{Z'} = (2 g_V^u + g_V^d)^2 s_X^2/4\pi 
\approx s_X^2 (4.2 \times 10^{-5})$.
Then we have the ratio
\be
\frac{\Gamma(\pi^0 \to \gamma Z')}{\Gamma(\pi^0 \to \gamma \gamma)}
 = \frac{2\,\alpha_{Z'}}{\alpha} 
     \left( 1- \frac{m^2_{Z'}}{m^2_{\pi^0}} \right)^3
 \approx s_X^2 \times 10^{-2} \times 
     \left( 1- \frac{m^2_{Z'}}{m^2_{\pi^0}} \right)^3 ,
\ee
where the factor 2 is the symmetric factor.
As is shown in Fig.~1,
$s_X$ is almost determined by $m_{Z'}$.
Hence we will obtain the number of produced $Z'$ 
from the number of $\pi^0$ production times above ratio
with respect to $m_{Z'}$. 

The next source of $Z'$ production is the $\eta$ meson decay
and the $\eta$ decay rate into $Z'$ is obtained by the ratio
\be
\frac{\Gamma(\eta \to Z' X)}
     {\Gamma(\eta \to \gamma X)}
 = \frac{ \alpha_{Z'}}{\alpha} 
   \left( 2\,{\rm Br}(\eta \to \gamma \gamma)
	 + 6\,{\rm Br}(\eta \to \pi^0 \pi^0 \pi^0)
 + 2\,{\rm Br}(\eta \to \pi^+ \pi^- \pi^0)
 + {\rm Br}(\eta \to \pi^+ \pi^- \gamma) \right)
    \left( 1- \frac{m^2_{Z'}}{m^2_{\eta}} \right)^3,
\ee
by considering the major decay channels of $\eta$ meson.
If $m_{Z'}>m_{\pi^0}$, 
the $\eta$ meson decay is
the dominant production mechanism of $Z'$  
and only the $\eta \to \gamma \gamma$ decay
contributes to the dark $Z$ production in Eq.~(25).

\subsection{LHC}

The LHC searches for LLPs are typically targeted 
to the signal processes of massive particles
with mean proper lifetimes up to ${\cal O}(10)$ ns.
The ATLAS inner tracker acceptance 
for transverse decay position is 4 mm -- 300 mm, 
and for longitudinal distance $< 300$ mm.
Then we take the decay length cut as 4 mm - 424 mm.
The CMS selects events 
with transverse impact parameter values
between 0.01 cm and 10 cm.
We set $l_i=1$ mm and $l_f=424$ mm for the LHC search
considering both of the ATLAS and CMS detectors.

We take the dark $Z$ boson energy to be 1 GeV
and 10 GeV as examples instead of full simulation.
We can see that these choices of energies are enough 
for the prospect to discover the dark $Z$ in the future.
The LHC will deliver the integrated luminosity
of ${\cal L}_{\rm Tot}=150$ fb$^{-1}$ in the run 3 period.
The total number of produced $\pi^0$ and $\eta$ mesons 
can be read out from Fig.~3 of Ref.~\cite{Feng:2017uoz}
for the pion energy 0.01 -- 10$^4$ GeV
and the meson angle $10^{-5}$ -- $\pi/2$.
For the ATLAS and CMS detector, 
the inner tracker covers
the pseudorapidity range $\eta < 2.5$
corresponding to $\theta > 0.16$
\cite{ATLAS:2022izj,CMS:2020krr}.
In the region of $\theta > 0.16$ 
we estimate the numbers of $\pi^0$
$N_{\pi^0}^{\rm tot} \sim 2 \times 10^{16}$ 
when $E_{\pi^0} \sim p_{\pi^0} = 2$ GeV,
$N_{\pi^0}^{\rm tot} \sim 1.5 \times 10^{14}$ 
when $E_{\pi^0} \sim p_{\pi^0} = 20$ GeV,
and the numbers of $\eta$
$N_{\eta}^{\rm tot} \sim 2 \times 10^{15}$ 
when $E_{\eta} \sim p_{\eta} = 2$ GeV,
$N_{\eta}^{\rm tot} \sim 2 \times 10^{13}$ 
when $E_{\eta} \sim p_{\eta} = 20$ GeV,
in Ref.~\cite{Feng:2017uoz}.

The expected number of the displaced vertices events
is obtained by
\be
N_{\rm events} = 
		 \left( N_{\pi^0}^{\rm tot} ~
                 \frac{\Gamma(\pi^0 \to \gamma Z')}
                 {\Gamma(\pi^0 \to \gamma \gamma)} 
		 + N_{\eta}^{\rm tot} ~
                 \frac{\Gamma(\eta \to Z'X)}
		 {\Gamma(\eta \to \gamma X)} \right)
		 \cdot P_{\rm decay}
		 \cdot \epsilon_{f},
\ee
where $ \epsilon_{f}$ is the detection efficiency
for the selection of tracks and the reconstruction of vertices.
We assume that $\epsilon_f = 10$ \% in this work 
from the recent study of the ATLAS group 
\cite{Wakida:2022swd},
where the efficiency $> 10$ \% is reported
for the long-lived supersymmetric particle searches
and the vertex points located inside the radius $\sim 25$ mm,
although those processes are different from ours.

We show the LHC sensitivity region
for the $E_{Z'}=1$ GeV and $E_{Z'}=10$ GeV cases
on the dark $Z$ boson mass and mixing plane,
$(m_{Z'},|\sin \theta_X|)$, 
together with the allowed parameters 
by the EW and Higgs sector constraints
in Fig.~3.
We demand that $N_{\rm events} > 25$ for the 5$\sigma$ discovery
of the $Z' \to e^- e^+$ events.
We can see that the LHC searches for the dark $Z$ cover 
the mass range $10 < m_{Z'}<150~{\rm MeV}$ in the LHC run 3.

\subsection{SHiP}

The SHiP is a newly proposed fixed target experiment 
of general purpose at the CERN SPS
to search for LLPs directly.
The high intensive proton beam of the SPS
with a dense target may provide the large number of 
light hadrons, e.g. $\pi^0$ and $\eta$,
which can decay into the dark $Z$ boson.
The produced dark $Z$ boson could decay 
within an isolated fiducial volume of the vacuum vessel.
The SHiP detector is located downstream of the decay volume
and detects the decay products of the dark $Z$ boson
by reconstructing the decay vertex and the mass of the $Z'$.
Due to the long decay volume,
it will provide a large lifetime acceptance.

\begin{figure}[t]
\centering
\includegraphics[width=14cm]{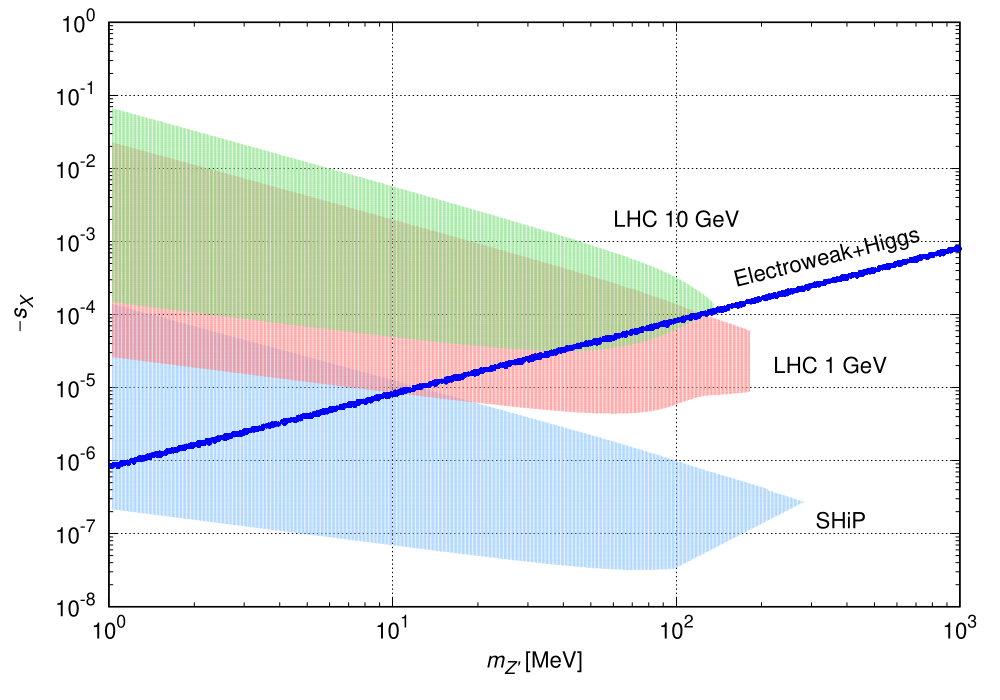}
\caption{
The sensitivity of the LHC and SHiP experiments 
on $(m_{Z'}, |\sin \theta_X|)$ plane.
The upper region for the LHC is obtained
under the assumption that $E_{Z'}=10$ GeV,
while the lower region for the LHC
under the assumption that $E_{Z'}=1$ GeV.
The blue line denotes the allowed parameters
given in Fig.~1.
}
\end{figure}

The number of events of $Z'$ decay inside the vacuum vessel 
is computed as
\be
N_{\rm events} = N_{PoT} \cdot 
                 \left( N_{\pi^0} ~
                 \frac{\Gamma(\pi^0 \to \gamma Z')}
                 {\Gamma(\pi^0 \to \gamma \gamma)} 
                 + N_{\eta} ~
                 \frac{\Gamma(\eta \to Z'X)}
		 {\Gamma(\eta \to \gamma X)} \right)
		 \cdot {\cal A}_{g} \cdot P_{\rm decay} \cdot 
		 {\rm Br}(Z' \to e^- e^+) \cdot \epsilon_{\rm det},
\ee
where $N_{PoT}$ is the number of protons on target and 
expected to be $6 \times 10^{20}$ for 15 years run
\cite{SHiP-2023proposal}.
The numbers of $\pi^0$ and $\eta$ productions per PoT
as $N_{\pi^0}\approx 42$ and $N_{\eta}\approx 5.5$ 
are estimated with FairShip software
\cite{SHiP:2020noy}.
The geometric acceptance ${\cal A}_g$ denotes the fraction
of the dark $Z$ inside the angular coverage of the vacuum vessel
from the proton beam dump and
${\cal A}_g \sim 30 \%$ is assumed
\cite{SHiP:2020noy}.
The detection efficiency $\epsilon_{\rm det}$ includes
the reconstruction efficiency and the selection efficiency
and is taken to be $\epsilon_{\rm det} = 50 \%$
\cite{SHiP:2020vbd}.
In the SHiP experiment, 
$l_i$ is the distance from the target 
to the entrance of the vacuum vessel, and
$l_f$ the distance from the target 
to the end of the vacuum vessel.
The designed values of the SHiP are
$l_i = 37$ m and $l_f = 87$ m at present
\cite{SHiP-2023proposal}

Figure 3 also depicts the SHiP sensitivity region
on the $(m_{Z'},|\sin \theta_X|)$ plane
with $N_{\rm events} > 25$ condition.
We can see that the SHiP will investigate
$2 m_e < m_{Z'} < 15~{\rm MeV}$ range in 15 years run.
The dark $Z$ boson is assumed to be produced
from only the $\pi^0$ and $\eta$ meson decay,
and have probed the mass range $m_{Z'} < m_{\eta}$.
We find that the higher mass region 
will not be probed by these experiments from this plot.
We foresee that no more allowed points will be examined 
if the dark $Z$ has more energy $>10$ GeV. 
On the other hand, the lower energy $Z'$ coverage at the LHC
overlaps with the region probed by the SHiP.
In conclusion,
we expect that the region $2 m_e < m_{Z'} < 300$ MeV 
can be probed by combining the LHC run 3 data
and the future SHiP experiment.

\section{Dark Matter Phenomenology}

In this section,
we show that the the singlet fermion in the hidden sector 
can be a DM candidate
with the dark $Z$ boson considered in the previous sections.
The dark $Z$ boson mass is assumed that 
$2 < m_{Z'} < 200$ MeV in this section.
The U(1)$_X$ charge $g_X X$ and the mass $m_X$
of the singlet fermion are new parameters 
of the hidden sector.
They are essentially free parameters, but
the perturbativity of coupling $|g_X X|^2 \le 4 \pi$
and kinematic condition $m_X > m_{Z'}/2$ are assigned.
Other model parameter values of 
$ \sin \alpha, \tan \beta, m_{h}, m_\pm $
are those allowed by the constraints in Sec.~2.

\begin{figure}[t]
\centering
\includegraphics[width=8cm]{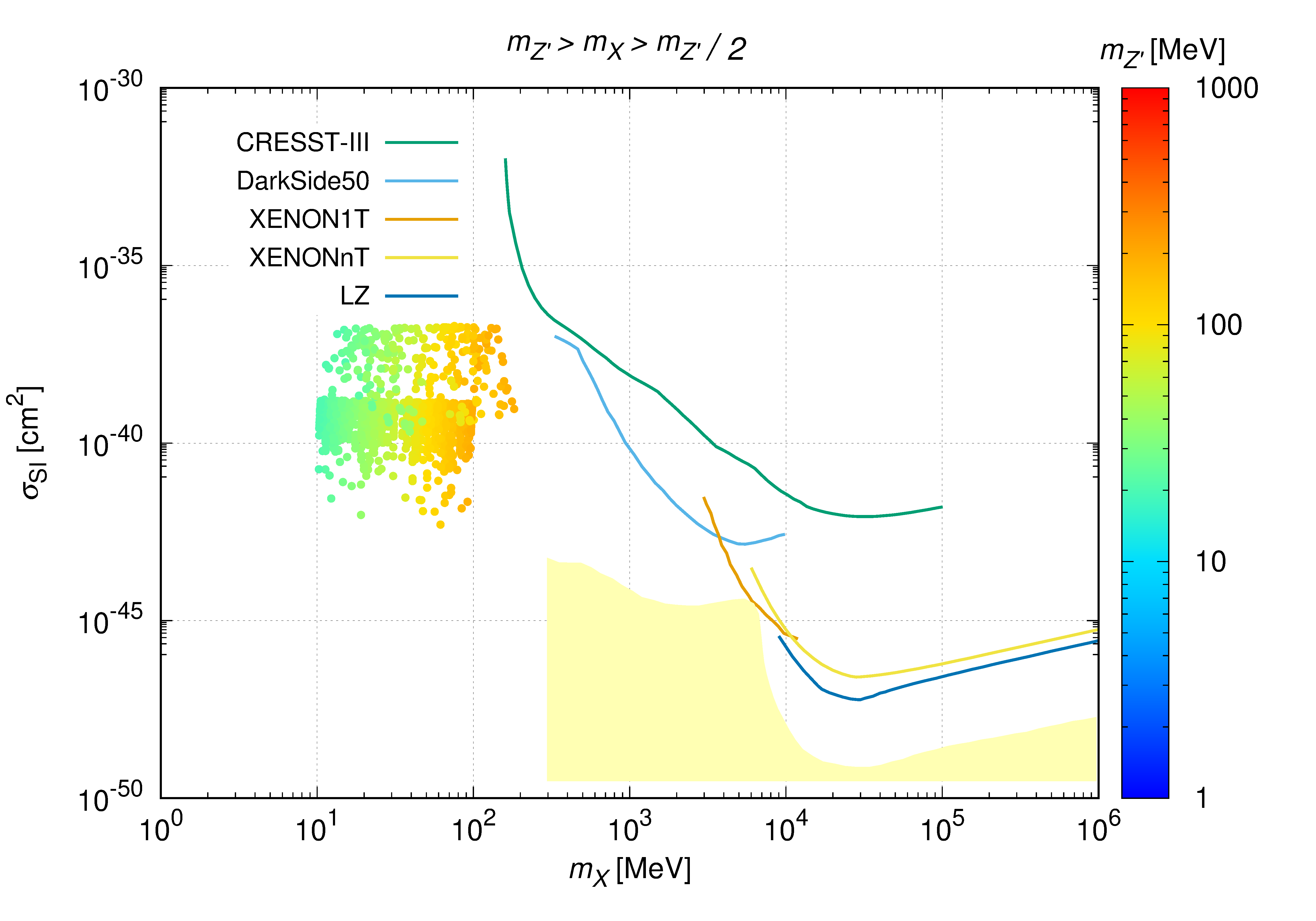}
\includegraphics[width=8cm]{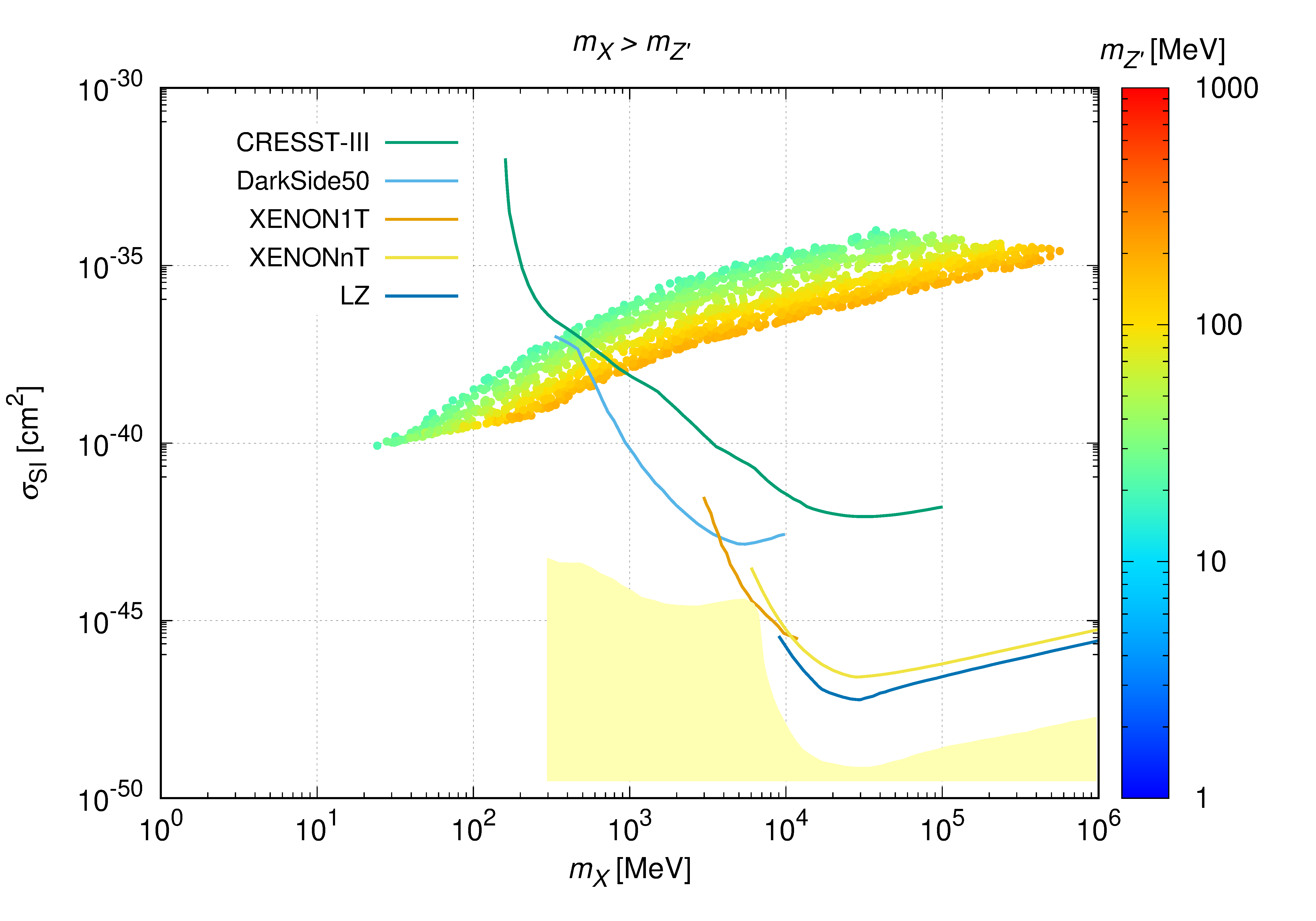}
\caption{
The DM-nucleon cross sections for the long-lived dark $Z$ boson
observable at the LHC and SHiP experiments.
}
\label{fig:DM}
\end{figure}

We compute the relic abundance and the DM-nucleon cross section 
with the public code \texttt{micrOMEGAs}
\cite{micromegas}. 
The main contribution to the relic density
comes from the DM annihilation processes,
$\psi_X \bar{\psi}_X \rightarrow Z' \rightarrow$ SM particles, 
$\psi_X \bar{\psi}_X \rightarrow Z' Z'$ 
and the Higgsstrahlung $\psi_X \bar{\psi}_X \rightarrow Z' h$ 
depending on the mass spectra.
Thermal freeze-out scenario is assumed
for the relic abundance calculation.
The most recent measurement of the DM contribution 
to the relic abundance $\Omega$ is given by
$\Omega_{\rm CDM} h^2 = 0.120 \pm 0.001$
from measurements of the anisotropy 
of the cosmic microwave background (CMB) 
and of the spatial distribution of galaxies
\cite{Planck:2018vyg}. 

Figure~\ref{fig:DM} depicts the DM-nucleon cross sections 
with the parameter sets consistent with the observed relic density 
within 3$\sigma$ range. 
We also impose the bullet cluster constraint, 
$\sigma / m_{DM} \lesssim 1 \mathrm{cm^2/g}$ 
\cite{Randall:2008ppe} 
to exclude the parameter set which yields too large DM 
self interaction. 

The left panel in Fig.~\ref{fig:DM} 
shows the DM-nucleon cross sections
for $m_{Z'}/2 < m_X < m_{Z'}$. 
The experimental bounds from CRESST III \cite{CRESST:2019jnq}, 
XENON1T \cite{XENON:2020gfr}, 
XENONnT \cite{XENON:2023cxc},
DarkSide-50 \cite{DarkSide:2022dhx} 
and LZ \cite{LZ:2022lsv} are shown together. 
In this case, 
the DM annihilation processes,
$\psi_X \bar{\psi}_X \rightarrow Z' \rightarrow$ SM particles, 
and the Higgsstrahlung $\psi_X \bar{\psi}_X \rightarrow Z' h$ 
contributes to the relic density.
Especially the DM pair annihilation 
$\psi_X \bar{\psi}_X\rightarrow Z' \rightarrow$ SM SM 
is enhanced through the resonance 
when $2\,m_{X} \sim m_{Z'}$. 
This enhancement constrain the couplings
to fit the observed relic density,  
which makes the DM-nucleon cross section smaller. 

The right panel in Fig.~\ref{fig:DM} denotes the case 
that the DM fermion is heavier than the dark $Z$, $m_X > m_{Z'}$. 
In this case, the $t$-channel, 
$\psi_X \bar{\psi}_X \rightarrow Z' Z'$ opens and dominates.
We find that
the DM mass is limited below  $\sim 200$ MeV
by exclusion with the direct detection experiments.

\section{Concluding Remarks \label{conclusion}}

We have studied the lifetime of the dark $Z$ boson 
which is the mediator in the fermionic DM model.
Because of the small coupling of the dark $Z$ boson,
its lifetime may be long, up to the microsecond, 
and experimental signatures can appear
in the present and future experiments.
We have explored the possibility to discover the displaced decays
inside the ATLAS, CMS and SHiP detectors.
Although We have not performed the full simulation,
we can obtain the sensitivity of the LHC run 3 
and the future SPS beam dump experiment
at a conservative estimate.
Combining the ATLAS and CMS at the LHC and the SHiP experiment,
we expect to find the displaced vertex 
of $Z' \to e^- e^+$ decays inside the detector in the near future
for the mass of $Z'$ boson, $2 m_e < m_{Z'} < 150$ MeV.

\acknowledgments
This work is supported 
by Basic Science Research Program
through the National Research Foundation of Korea (NRF)
funded by the Ministry of Science, ICT, and Future Planning 
under the Grant No. NRF-2021R1A2C2011003 (DWJ, KYL, CY), 
No. NRF-2020R1I1A1A01073770  and RS-2023-00247615 (CY).
The work of DWJ is supported in part by 
NRF-2019R1A2C1089334, NRF-2021R1A2B5B02087078, RS-2023-00246268 
and the Yonsei University Research Fund 
(Post Doc. Researcher Supporting Program) of 2022 (project No.: 2022-12-0035).

\def\PRDD #1 #2 #3 {Phys. Rev. D {\bf#1},\ #2 (#3)}
\def\PRD #1 #2 #3 #4 {Phys. Rev. D {\bf#1},\ No. #2, #3 (#4)}
\def\PRLL #1 #2 #3 {Phys. Rev. Lett. {\bf#1},\ #2 (#3)}
\def\PRL #1 #2 #3 #4 {Phys. Rev. Lett. {\bf#1},\ No. #2, #3 (#4)}
\def\PLB #1 #2 #3 {Phys. Lett. B {\bf#1},\ #2 (#3)}
\def\NPB #1 #2 #3 {Nucl. Phys. B {\bf #1},\ #2 (#3)}
\def\ZPC #1 #2 #3 {Z. Phys. C {\bf#1},\ #2 (#3)}
\def\EPJ #1 #2 #3 {Euro. Phys. J. C {\bf#1},\ #2 (#3)}
\def\JPG #1 #2 #3 {J. Phys. G: Nucl. Part. Phys. {\bf#1},\ #2 (#3)}
\def\JHEP #1 #2 #3 {JHEP {\bf#1},\ #2 (#3)}
\def\JCAP #1 #2 #3 {JCAP {\bf#1},\ #2 (#3)}
\def\IJMP #1 #2 #3 {Int. J. Mod. Phys. A {\bf#1},\ #2 (#3)}
\def\MPL #1 #2 #3 {Mod. Phys. Lett. A {\bf#1},\ #2 (#3)}
\def\PTP #1 #2 #3 {Prog. Theor. Phys. {\bf#1},\ #2 (#3)}
\def\PR #1 #2 #3 {Phys. Rep. {\bf#1},\ #2 (#3)}
\def\RMP #1 #2 #3 {Rev. Mod. Phys. {\bf#1},\ #2 (#3)}
\def\PRold #1 #2 #3 {Phys. Rev. {\bf#1},\ #2 (#3)}
\def\IBID #1 #2 #3 {{\it ibid.} {\bf#1},\ #2 (#3)}

\end{document}